\documentclass[fp,twocolumn]{jpsj3}
\usepackage{txfonts}
\usepackage{braket}
\usepackage{color}

\title{Microwave Scattering in the Subohmic Spin-Boson Systems of Superconducting Circuits}

\author{Tsuyoshi Yamamoto\thanks{t.yamamoto@issp.u-tokyo.ac.jp} and Takeo Kato}
\inst{Institute for Solid State Physics, The University of Tokyo, Kashiwa, Chiba 277-8581, Japan} 

\abst{
We study quantum critical phenomena in the microwave scattering of the subohmic spin-boson system, which exhibits a quantum phase transition at a critical system-reservoir coupling.
By relating the reflection coefficient of a microwave with the dynamic susceptibility of the subohmic spin-boson system, we clarify the appearance of quantum critical phenomena in microwave scattering.
Further, we propose experimental setups to realize the subohmic spin-boson system in a superconducting circuit composed of a charge qubit and a dissipative transmission line.
}

\begin{document}
\maketitle

\section{Introduction}
\label{sec.Intro}

The ubiquity of quantum critical phenomena (QCP) in strongly correlated systems has been investigated in a number of studies~\cite{Sachdev2011}.
Nonetheless, it remains challenging to observe QCP in well-controllable systems by the fine adjustment of experimental parameters.
Recent developments in nanostructure fabrication and measurement have enabled us to design nanoscale devices for the study of novel quantum phenomena in a system strongly interacting with an environment.
For example, QCP in the two-channel Kondo model have been studied in detail in quantum dot systems, as effectively described by a local spin interacting antiferromagnetically with an electronic environment~\cite{Potok2007,Mebrahtu2012,Mebrahtu2013,Keller2015,Iftikhar2015,Cox1998,Vojta2006,Bulla2008}.
It is natural to ask whether one can realize a counterpart in controllable bosonic systems that display quantum phase transitions (QPTs).

The spin-boson model has been studied as a minimal model of a quantum two-state system interacting with a bosonic environment for a long time~\cite{Leggett1987,Weiss1999}.
This model has a variety of applications, including superconducting circuits~\cite{Peropadre2013,Leppakangas2018}, photon waveguides~\cite{Bourassa2009}, and molecular junctions~\cite{Nitzan2006,Ren2010}.
The properties of the bosonic reservoir are characterized by a spectral density, $I(\omega)\propto\omega^{s}$.
On the one hand, the ohmic spin-boson system ($s=1$) has been studied in various contexts in a great deal of theoretical work~\cite{Segal2010,Ruokola2011,LeHur2012,Saito2013,Segal2014,Taylor2015,Wang2015,Agarwalla2017,Leppakangas2018,Forn-Diaz2017,Magazzu2018}, because it displays remarkable phenomena such as the Kosterlitz-Thouless-type phase transition~\cite{Leggett1987,Anderson1971,Kosterlitz1976} and the Kondo effect~\cite{Leggett1987,Anderson1971,Guinea1985,Saito2013}.
On the other hand, the subohmic spin-boson system ($s<1$) induces a second-order phase transition at zero temperature by tuning the system-reservoir coupling to a critical value~\cite{Kehrein1996a,Kehrein1996b,Weiss1999,Bulla2003,Vojta2005,Vojta2009,Anders2007,LeHur2007,Lu2007,Winter2009,Zhang2010,Florens2011,Vojta2012,Tong2012,Yamamoto2018_QCP}.
The present authors studied QCP under heat transport through evaluation of the temperature dependence of thermal conductance~\cite{Yamamoto2018_QCP}.
Nevertheless, studying QCP using standard techniques of measurement would be preferable, given the expected difficulty of measuring thermal conductance.

Experimental techniques on light-matter interaction systems have greatly improved over the past few decades~\cite{Kockum2019,Martinez2019}.
Recently, the ohmic spin-boson model has been realized in a flux qubit coupled to an $LC$ transmission line, where experimental microwave scattering measurements have been performed~\cite{Forn-Diaz2017,Magazzu2018}.
This development suggests the possibility of probing the QCP in the subohmic spin-boson system by microwave scattering in a transmission line weakly coupled to it.
Realization of the subohmic reservoir in a circuit model was first discussed for the special case of $s=1/2$ in Ref.~\citen{Tong2006} and, recently, for arbitrary values of $s$ in the range $0<s<1$ in Ref.~\citen{Yamamoto2018_QCP} by the present authors.
In the latter paper, a flux qubit coupled to an $RLC$ series circuit was effectively described by the subohmic spin-boson model.

In this paper, QCP are studied in a charge qubit coupled to the subohmic reservoir, and microwave scattering is considered with an additional transmission line weakly coupled to the charge qubit.
The input-output theory~\cite{Clerk2010,LeHur2012,Goldstein2013} is used to derive the relation between the reflection coefficient and the dynamic susceptibility of the subohmic spin-boson system.
Furthermore, this study clarified the appearance of quantum critical behavior in the frequency dependence of the dynamic susceptibility by analytic discussion as well as numerical calculations based on the continuous-time quantum Monte Carlo (CTQMC) method.
Thus, the dynamic susceptibility in the quantum critical regime was confirmed to exhibit distinctive power-law frequency dependence reflected by the nature of QPTs.
Three experimental setups are proposed to realize the subohmic reservoir using the charge qubit and the $RLC$ transmission line.
One of the proposed setups has the advantage of the number of circuit elements being less than that proposed previously~\cite{Tong2006,Yamamoto2018_QCP}.

The rest of this paper is organized as follows.
The model, which consists of a charge qubit, a subohmic reservoir, and a transmission line, is introduced in Sect.~\ref{sec:model}.
In Sect.~\ref{sec:input-output}, the reflection coefficient of microwave scattering in a transmission line is related to the dynamic susceptibility of the subohmic spin-boson system.
A brief summary of the analytic results on the dynamic susceptibility is provided in Sect.~\ref{sec:dynamic_sus}, followed by numerical evaluation as well as the calculation of the reflection by the CTQMC method in Sect.~\ref{sec:result}.
Circuit models realizing the subohmic spin-boson system are proposed in Sect.~\ref{sec:realization}.
A summary of the results is given in Sect.~\ref{sec:summary}.

\section{Model}
\label{sec:model}

\begin{figure}[tbp]
	\centering
	\includegraphics[width=8cm]{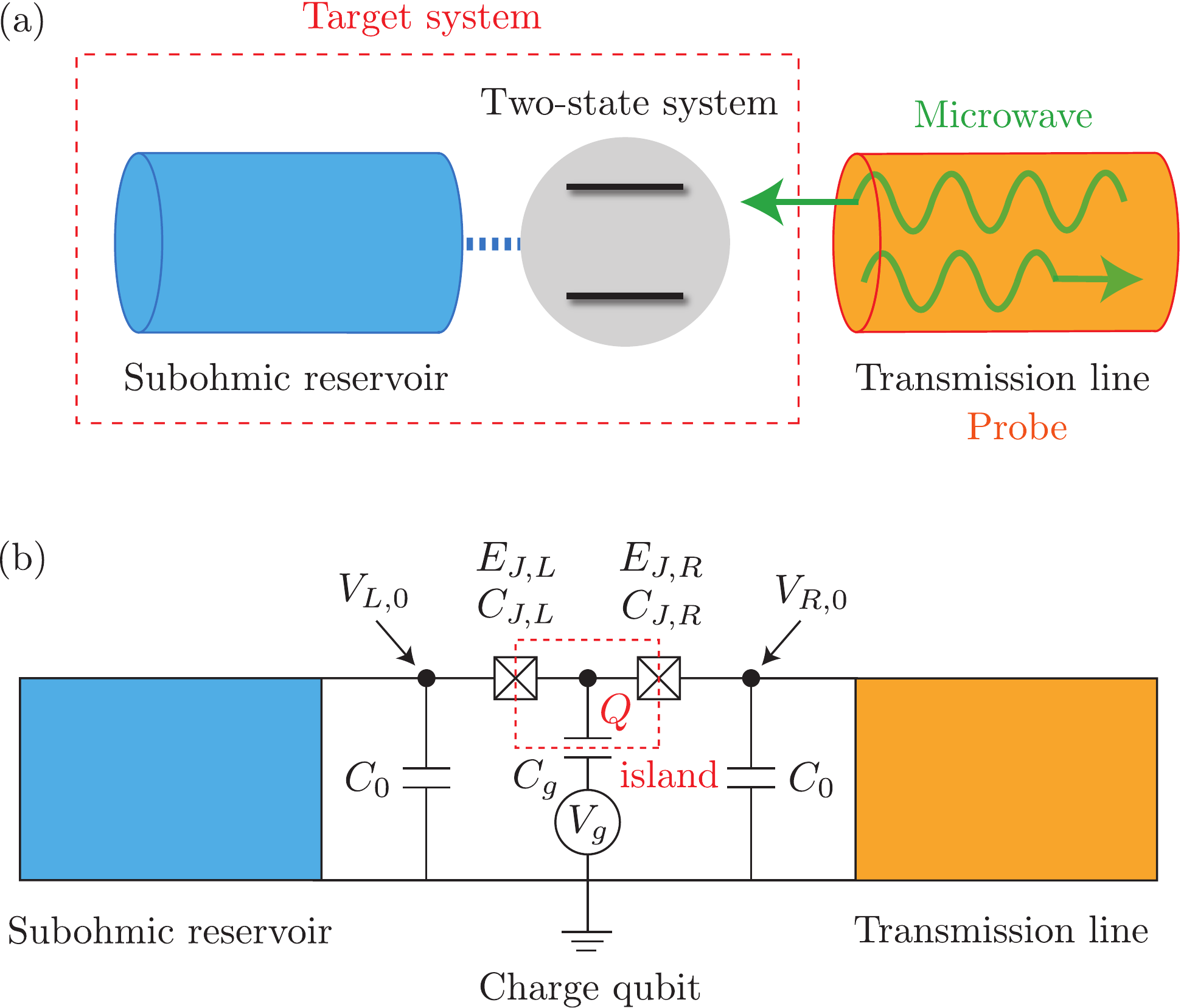}
	\caption{
    (Color online) (a) Schematic representation of the spin-boson model (the target system), which comprises a two-state system coupled to a subohmic reservoir.
    A transmission line weakly coupled to the target system is introduced as a probe.
    The microwave is injected from the transmission line into the target system.
    (b) Superconducting circuit for a charge qubit coupled to the subohmic reservoir and the transmission line.
    }
	\label{fig:model}
\end{figure}

We prepare a target system, consisting of a two-state system and a subohmic reservoir, and consider microwave scattering in the target system within the linear response theory (Fig.~\ref{fig:model}~(a)).
An incident microwave enters from a transmission line into the charge qubit, and a reflected microwave is allowed to propagate in the same transmission line in the opposite direction.
The frequency dependence of the microwave loss at the target system can be used to probe the system's quantum critical behavior, that is, the subohmic spin-boson system.

\subsection{Charge qubit}

We consider a superconducting circuit shown in Fig.~\ref{fig:model}~(b) for a charge qubit composed of two Josephson junctions
with Josephson energies $E_{J,\nu}$ and capacitances $C_{J,\nu}$ ($\nu = L,R$).
The charge state of the qubit is controlled by a gate voltage $V_g$ via a gate capacitance $C_g$.
If no excitation occurred in both the subohmic reservoir and the transmission line, their output voltages, $V_{L,0}$ and $V_{R,0}$, respectively, are equal to zero.
Thus, the Hamiltonian of the charge qubit can be expressed as
\begin{align}
    \label{eq:H_circuit}
    H_{\rm S} &= E_{C}(n-n_{g})^2 - \sum_{\nu=L,R} E_{J,\nu}\cos(\phi_{\nu}/\phi_{0}),
\end{align}
where $E_{C} = 2e^2/C_{\Sigma}$ ($C_{\Sigma} = C_{g}+C_{J,L}+C_{J,R}$: total capacitance) is the Coulomb energy, $n_g$ is the offset Cooper-pair number defined by $n_g \equiv Q_{g}/2e$ ($Q_g$: charge accumulated at the gate capacitance), $\phi_{\nu}$ is the phase difference across the Josephson junction, and $\phi_{0}=h/2e$ is a flux quantum.
The operator $n$ describes the number of excess Cooper pairs at the superconducting island ($Q = 2en$).
Considering small Josephson junctions ($E_{J,L},~E_{J,R} \ll E_{C}$) and setting the offset charge as $n_{g} =0.5$, we can effectively describe the system by only the two charge states ($\ket{n=0}$ and $\ket{n=1}$), whose charging energies are well separated from the others.
After truncating the system into a two-state system, the Hamiltonian can be written as
\begin{align}
H_{\rm S} &= -\frac{\hbar\Delta}{2}\sigma_{x} -\varepsilon\sigma_z,
\end{align}
where the first term describes the Josephson energy, and the second term corresponds to detuning from the degenerate gate voltage $n_g=0.5$.
Here, $\Delta \equiv E_{\rm J}/\hbar$ is the tunneling frequency and $\sigma_i$ ($i=x,y,z$) is the Pauli matrix.
We rewrote the two charge states, $\ket{n=1}$ and $\ket{n=0}$, with $\ket{\sigma_z=\pm 1}$ and replaced the number operator $n$ with $\sigma_z = 2n-1$.
In this paper, we consider the symmetric case ($\varepsilon = 0$) and use only the detuning energy to define the static susceptibility.

\subsection{Subohmic reservoir}
\label{sec:circuit_model}

The Hamiltonian of the target system (the charge qubit coupled to the subohmic reservoir) can be expressed as
\begin{align}
H_{\rm sub} = H_{\rm S} + H_{{\rm R},{\rm sub}} + H_{{\rm I},{\rm sub}}.
\end{align}
The second and third terms on the right-hand side describe the reservoir and the system-reservoir interaction, respectively, which are generally expressed as
\begin{align}
    H_{{\rm R},{\rm sub}} &= \sum_{k}\hbar\omega_k b_k^{\dagger} b_k,
    \label{eq:H_sb} \\
    H_{{\rm I},{\rm sub}} &= -\frac{\sigma_z}{2}\sum_{k}\hbar\lambda_k \left(b_k + b_k^{\dagger}\right),
    \label{eq:H_sb_I}
\end{align}
respectively.
This Hamiltonian is called the spin-boson Hamiltonian.
Here, $b_k$ ($b_k^{\dagger}$) is an annihilation (creation) operator of excitation with a frequency $\omega_k$ and a system-reservoir coupling constant $\lambda_k$ in the subohmic reservoir.
The properties of the reservoir are characterized by the spectral density
\begin{align}
    I_{\rm sub}(\omega) \equiv \sum_{k}\lambda_k^2\delta(\omega-\omega_k).
    \label{eq:spectral_def}
\end{align}
In the numerical calculation shown in Sect.~\ref{sec:result}, we assume that the spectral density takes a form
\begin{align}
    I_{\rm sub}(\omega) = 2\alpha \omega \left(\frac{\omega}{\omega_{\rm c}}\right)^{s-1} e^{-\omega/\omega_{\rm c}},
\end{align}
for simplicity, where $\alpha$ is the dimensionless system-reservoir coupling strength, and $\omega_{\rm c}$ is the cutoff frequency.
The exponent of spectral density can be taken as $s < 1$ (the subohmic reservoir).

The two-state system coupled to the subohmic reservoir displays a QPT when the coupling strength is tuned to the critical point $\alpha_{\rm c}$, which is a function of $s$ and $\Delta/\omega_{\rm c}$~\cite{Kehrein1996a,Kehrein1996b,Weiss1999,Winter2009,Yamamoto2018,Yamamoto2018_QCP}.
For $\alpha < \alpha_{\rm c}$, the ground state is expressed by the coherent superposition of two $\sigma_z$-basis states, $\Ket{\sigma_z = \pm 1}$.
This phase is called the ``delocalized phase''.
For $\alpha > \alpha_{\rm c}$, the ground state becomes twofold degenerate with a completely broken quantum coherence. 
This phase is called the ``localized phase''.

Explicit construction of the subohmic reservoir using an $RLC$ transmission line is discussed in Sect.~\ref{sec:realization}.
There, the interaction between the charge qubit and the subohmic reservoir can be expressed as
\begin{align}
 H_{{\rm I},{\rm sub}} &= -\frac{|e|C_{J,L}}{C_{\Sigma}}\sigma_{z}V_{L,0},
  \label{eq:H_sb_I2} \\
    V_{L,0} &= \frac{C_{\Sigma}}{2|e|C_{J,L}}
    \sum_{k}\hbar\lambda_k\left(b_k + b_k^{\dagger}\right)
    \label{eq:V_L} ,
\end{align}
where $V_{L,0}$ is the output voltage of the subohmic reservoir (Fig.~\ref{fig:model}~(b)).
We introduce the retarded voltage-voltage correlation function defined by
\begin{align}
    G_{V}^{\rm R}(t) \equiv -\frac{i}{\hbar}\theta(t)\Braket{\left[V_{L,0}(t),V_{L,0}(0)\right]}_0,
\end{align}
where $\theta(t)$ is the Heaviside step function, $V_{L,0}(t) = e^{iH_{{\rm R},{\rm sub}}t/\hbar} V_{L} e^{-iH_{{\rm R},{\rm sub}}t/\hbar}$, and $\Braket{\cdots}_0$ indicates an ensemble average with respect to $H_{{\rm R},{\rm sub}}$.
Using Eq.~(\ref{eq:spectral_def}), the imaginary part of the Fourier transformation of $G_{V}^{\rm R}(t)$ can be expressed in terms of the spectral density:
\begin{align}
    \label{eq:G_I}
    {\rm Im}\left[G_{V}^{\rm R}(\omega)\right] = -\pi\hbar\left(\frac{C_{\Sigma}}{2|e|C_{J,L}}\right)^{2}I_{\rm sub}(\omega).
\end{align}
Using the linear response theory~\cite{Bruus2004}, the voltage-voltage correlation function $G_{V}^{\rm R}(t)$ is related with the impedance $Z_{\rm sub}(\omega)$ of the subohmic reservoir as
\begin{align}
    \label{eq:G_linear_response}
    G_{V}^{\rm R}(\omega) = -i\omega Z_{\rm sub}(\omega).
\end{align}
Therefore, the spectral function can be related with the impedance of the circuit as
\begin{align}
I_{\rm sub}(\omega) = \frac{1}{\pi \hbar}
\left(\frac{2|e|C_{J,L}}{C_{\Sigma}}\right)^{2} \omega {\rm Re}\left[ Z_{\rm sub}(\omega)\right] .
\label{eq:ImpedanceFormula}
\end{align}

\subsection{Transmission line}
\label{subsec:LC_transmission}

\begin{figure}[tbp]
	\centering
	\includegraphics[width=8cm]{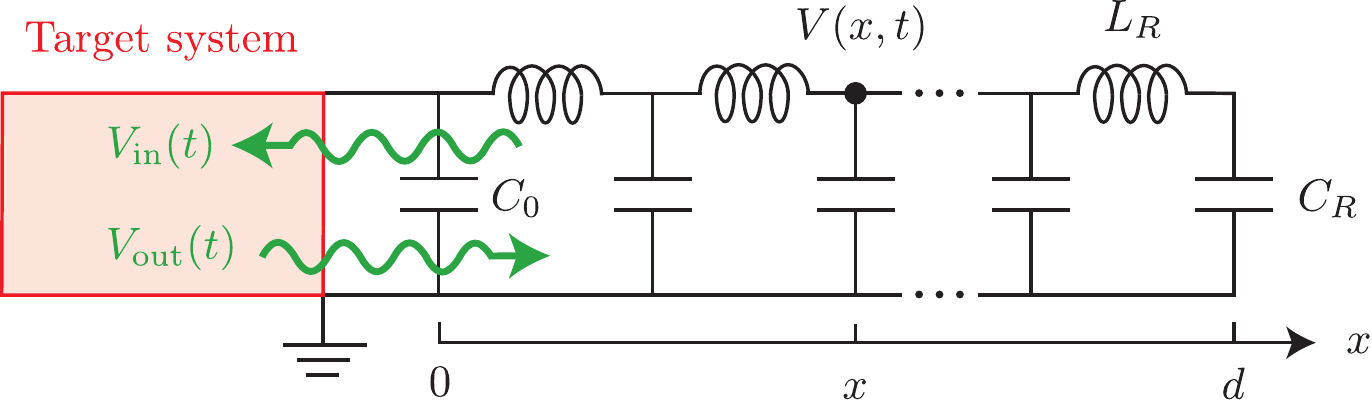}
	\caption{
    (Color online) The $LC$ transmission line coupled to the
    target system.
    Input mode $V_{\rm in}(t)$ is injected into the target system from the $LC$ transmission line.
    Output mode $V_{\rm out}(t)$ is radiated into the $LC$ transmission line from the target system.
    The target system represents a two-state system coupled to the subohmic reservoir (see Fig.~\ref{fig:model}).
    }
	\label{fig:input-output}
\end{figure}

We confirm critical behavior near the QPT after preparing a transmission line weakly coupled to the target system (Fig~\ref{fig:model}).
The transmission line is described by a continuum limit of the $LC$ circuit in Fig.~\ref{fig:input-output}.
The phase operator is defined as
\begin{align}
    \label{eq:flux_def}
    & \phi(x,t) \equiv \int_{-\infty}^{t}dt'~V(x,t'),
\end{align}
where $V(x,t)$ is the voltage on the $LC$ transmission line at position $x$ at time $t$.
Classical partial differential equations for the phase operator $\phi(x,t)$ can be derived from the Hamiltonian defined by~\cite{Blais2004,Peropadre2013}
\begin{align}
    \label{eq:H_transmission}
    H_{{\rm R},{\rm line}} = \int_{0}^{d}dx~\left\{\frac{1}{2c}q(x)^2 + \frac{1}{2l}\left[\partial_{x}\phi(x)\right]^2\right\},
\end{align}
where $c$ and $l$ are the capacitance and the inductance per unit length, respectively, $d$ is the length of the transmission line, and $q(x)$ is the charge density operator conjugate to $\phi(x)$.
For a description of the quantum dynamics, we impose an exchange relation
\begin{align}
\left[\phi(x),q(x')\right] = i\hbar\delta(x-x'). \label{eq:ExchangeRelation1}
\end{align}
Using a Bosonic operator
\begin{align}
    B_{k}
    \equiv \frac{1}{\sqrt{\hbar\Omega_{k}d}}\int_{0}^{d}dx~e^{-ikx}\left[\frac{q(x)}{\sqrt{2c}} - i\sqrt{\frac{k^2}{2l}}\phi(x)\right],
\end{align}
the Hamiltonian of the transmission line is diagonalized as
\begin{align}
    H_{{\rm R},{\rm line}} = \sum_k \hbar \Omega_k B_k^\dagger B_k,
\end{align}
where $\Omega_{k} = |k|/\sqrt{lc}$. 
Here, note that the commutation relation $\left[B_{k},B_{k'}^{\dagger}\right] = \delta_{k,k'}$ holds for Eq.~(\ref{eq:ExchangeRelation1}).

We consider the capacitive coupling between the charge qubit and the transmission line, with the voltage written by the field operators, $B_k$ and $B_k^{\dagger}$, as
\begin{align}
    \label{eq:voltage}
    V(x) = \sum_{k}\sqrt{\frac{\hbar\Omega_k}{dc}}\left(B_ke^{ikx} + B_k^{\dagger}e^{-ikx}\right), 
\end{align}
and the output voltage at the qubit is written as $V_{R,0}=V(x=0)$.
The Hamiltonian of the interaction between the charge qubit and the transmission line is given as
\begin{align}
     H_{{\rm I},{\rm line}} &= -\frac{|e|C_{J,R}}{C_{\Sigma}}\sigma_{z}V_{R,0}=-\frac{\sigma_z}{2}\sum_{k} \hbar \Lambda_k \left(B_k + B_k^{\dagger}\right), \\
    \Lambda_k &= \frac{2|e|C_{J,R}}{C_{\Sigma}} \sqrt{\frac{\Omega_k}{\hbar dc}}.
\end{align}
The spectral density for the transmission line is obtained as
\begin{align}
I_{\rm line}(\omega) &\equiv \sum_{k}\Lambda_k^2 \delta(\omega-\Omega_k)
\nonumber \\
&= \frac{1}{\pi\hbar}\sqrt{\frac{l}{c}} \left(\frac{2|e|C_{J,R}}{C_{\Sigma}}\right)^2 \omega \equiv \alpha_{\rm line} \omega.
\end{align}
in the limit $d\rightarrow \infty$.
Note that the transmission line plays the role of the ohmic reservoir as the spectral density is proportional to $\omega$.
This is consistent with Eq.~(\ref{eq:ImpedanceFormula}) because the impedance of the $LC$ transmission line is $Z_{\rm line}(\omega) = \sqrt{l/c}$.

\section{Input-output Theory}
\label{sec:input-output}

In this section, we relate the reflection coefficient in microwave scattering with the dynamic susceptibility of the target system by combining the input-output theory~\cite{Clerk2010,LeHur2012,Goldstein2013} and the linear response theory.
We begin with the total Hamiltonian
\begin{align}
H = H_{\rm sub} + H_{{\rm R},{\rm line}}+ H_{{\rm I},{\rm line}},
\end{align}
and consider the Heisenberg equation of motion for the annihilation operator on the transmission line,
\begin{align}
    \label{eq:Heisenberg}
    \dot{B}_k(t)
    = \frac{i}{\hbar}\left[H,B_k(t) \right]
    = -i\Omega_k B_k(t) + i\frac{\Lambda_k}{2}\sigma_z(t),
\end{align}
where $O(t) = e^{iHt/\hbar}Oe^{-iHt/\hbar}$.
By integrating the Heisenberg equation of motion~(\ref{eq:Heisenberg}), the solution of this equation is given by~\cite{LeHur2012}
\begin{align}
    \label{eq:b_input}
    B_{k}(t) &= e^{-i\Omega_{k}(t-t_{0})}B_{k}(t_{0})
    + i\frac{\Lambda_{k}}{2}\int_{t_{0}}^{t}dt'~e^{-i\Omega_{k}(t-t')}\sigma_z(t'),
\end{align}
where $t_{0}$ is a past time before any excitation mode reaches the two-state system.
The first term represents the free time evolution in the transmission line, and the second term describes the effect of the interaction between the charge qubit and the transmission line.
Similarly, we obtain an alternative solution to Eq.~(\ref{eq:Heisenberg}) as
\begin{align}
    \label{eq:b_output}
    B_{k}(t) &= e^{-i\Omega_{k}(t-t_{1})}B_{k}(t_{1})
    - i\frac{\Lambda_{k}}{2}\int_{t}^{t_{1}}dt'~e^{-i\Omega_{k}(t-t')}\sigma_z(t'),
\end{align}
where $t_{1}$ is a future time after the entire excitation mode leaves the two-state system.
The voltages of input and output modes are defined as
\begin{align}
    \label{eq:V_in}
    V_{\rm in}(t) &\equiv \sum_{k}\sqrt{\frac{\hbar\Omega_{k}}{dc}}\left[e^{-i\Omega_{k}(t-t_{0})}B_{k}(t_{0}) + {\rm h.c.}\right], \\
    \label{eq:V_out}
    V_{\rm out}(t) &\equiv \sum_{k}\sqrt{\frac{\hbar\Omega_{k}}{dc}}\left[e^{-i\Omega_{k}(t-t_{1})}B_{k}(t_{1}) + {\rm h.c.}\right],
\end{align}
respectively.
Fourier transformation of the voltages of the input and output modes gives
\begin{align}
    \label{eq:V_in-V_out}
    \Braket{V_{\rm out}(\omega)} = \Braket{V_{\rm in}(\omega)} + i\frac{\pi\hbar C_{\Sigma}}{2|e|C_{J,R}}I_{{\rm line}}(\omega)\Braket{\sigma_z(\omega)},
\end{align}
where $I_{\rm line}(\omega) \propto \omega$ is the spectral density of the transmission line, $\Braket{\cdots} = {\rm tr}[e^{-\beta H}\cdots ]/{\rm tr}[e^{-\beta H}]$, and $\beta = 1/k_{\rm B}T$.

Assuming weak coupling between the two-state system and the ohmic reservoir ($\alpha_{\rm line} \ll 1$), the population $\Braket{\sigma_z(\omega)}$ in Eq.~(\ref{eq:V_in-V_out}) can be calculated using the linear response theory as
\begin{align}
    \label{eq:sigma_z_linear_response}
    \Braket{\sigma_z(\omega)} = \frac{|e|C_{J,R}}{C_{\Sigma}}\chi_{\rm sub}(\omega)\Braket{V_{\rm in}(\omega)}.
\end{align}
Here $\chi_{\rm sub}(\omega)$ is the Fourier transformation of the dynamic susceptibility of the two-state system coupled to the subohmic reservoir defined by
\begin{align}
    \label{eq:dynamic_sus_def}
    \chi_{\rm sub}(t) \equiv \frac{i}{\hbar}\theta(t)\Braket{\left[\sigma_z(t),\sigma_z(0)\right]}_{\rm sub},
\end{align}
where $\Braket{\cdots}_{\rm sub} = {\rm tr}[e^{-\beta H_{\rm sub}}\cdots]$.
From Eqs.~(\ref{eq:V_in-V_out}) and (\ref{eq:sigma_z_linear_response}), we obtain the reflection coefficient as
\begin{align}
    \label{eq:r_def}
    r(\omega)
    \equiv \frac{\Braket{V_{\rm out}(\omega)}}{\Braket{V_{\rm in}(\omega)}}
    = 1+i\frac{\pi\hbar}{2}I_{\rm line}(\omega)\chi_{\rm sub}(\omega).
\end{align}

\section{Dynamic Susceptibility}
\label{sec:dynamic_sus}

In this section, we briefly summarize the features of the dynamic susceptibility $\chi_{\rm sub}(\omega)$ at the critical point ($\alpha = \alpha_{\rm c}$) and in the delocalized regime ($\alpha < \alpha_{\rm c}$).

\subsection{Quantum critical regime}
\label{sec:dynamic_sus_QCP}

When the system-reservoir coupling is tuned at QPT ($\alpha=\alpha_{\rm c}$), the dynamic susceptibility exhibits distinctive frequency dependence reflecting the nature of the QPT: 
\begin{align}
    \label{eq:dynamic_sus_QCP}
    \chi_{\rm sub}(\omega) \sim \omega^{-y},\quad (\alpha = \alpha_{\rm c}).
\end{align}
The critical exponent $y$ is related to the critical exponent $\eta$ of the imaginary-time spin-spin correlation function:
\begin{align}
C(\tau)& \equiv\Braket{\sigma_z(\tau)\sigma_z(0)}_{\rm sub} \sim\tau^{-\eta+1}, \\
    \label{eq:x-eta}
    y &= 2 - \eta,
\end{align}
where $\sigma_z(\tau) = e^{H_{\rm sub}\tau/\hbar}\sigma_{z}e^{-H_{\rm sub}\tau/\hbar}$.
For $0 < s \le 0.5$, the QPT belongs to the mean-field universality class~\cite{Fisher1972,Luijten1997,Luijten1997_thesis,Winter2009,Vojta2012,Yamamoto2018_QCP}, and we obtain $\eta = 2-s$, which leads to
\begin{align}
    \label{eq:dynamic_sus_exponent}
    y = s,\quad (0 < s \le 0.5),
\end{align}
In contrast, the critical exponent $y$ becomes a complex function of $s$ for $0.5 < s < 1$, because the QPT belongs to a nontrivial universal class~\cite{Luijten1997_thesis}.

\subsection{Delocalized regime}

If the system-reservoir coupling is set sufficiently below the critical value ($\alpha < \alpha_{\rm c}$), the ground state would become a coherent superposition of the two charge states.
At low temperatures, we can apply the generalized Shiba relation~\cite{Sassetti1990,Yamamoto2018}:
\begin{align}
    \label{eq:Shiba}
    \lim_{\omega\rightarrow+0}\frac{\hbar{\rm Im\left[\chi_{\rm sub}(\omega)\right]}}{\omega^{s}} = 2\pi\alpha\omega_{\rm c}^{1-s}\left(\frac{\hbar\chi_{0,{\rm sub}}}{2}\right)^2,
\end{align}
where the static susceptibility $\chi_{0,{\rm sub}}$ is defined as
\begin{align}
    \label{eq:static_sus_def}
    \chi_{0,{\rm sub}} \equiv \lim_{\varepsilon\rightarrow0}\frac{\Braket{\sigma_z}_{\rm sub}}{\varepsilon}.
\end{align}
Using this generalized Shiba relation, the low-frequency dynamic susceptibility is obtained as
\begin{align}
    \label{eq:dynamic_sus_co-tunneling}
    {\rm Im}\left[\chi_{\rm sub}(\omega)\right] \sim \omega^{s}.
\end{align}

\section{Numerical Calculation}
\label{sec:result}

We demonstrate the critical behavior of the dynamic susceptibility by performing the CTQMC simulation~\cite{Winter2009,Yamamoto2018}, in which the correlation function $C(\tau) = \langle \sigma_z(\tau) \sigma_z(0) \rangle_{\rm sub}$ can be evaluated numerically (for details, see Ref.~\citen{Yamamoto2018}).
We define the Fourier transformation of the imaginary-time spin-spin correlation function as
\begin{align}
C(i\omega_{n}) = \int_{0}^{\beta}d\tau~e^{i\omega_{n}\tau} C(\tau),
\end{align}
where $\omega_{n} = 2\pi n/\hbar\beta$ is the Matsubara frequency.
The Monte Carlo data presented below represent averages over $10^{5}$ to $10^{6}$ cluster updates.
The dynamic susceptibility is obtained from $C(i\omega_{n})$ by the analytic continuation
\begin{align}
    \label{eq:analytic_continuation}
    \chi_{\rm sub}(\omega) = C(i\omega_{n}\rightarrow\omega+i\delta).
\end{align}
We use the Pad\'{e} approximation as continuation of the numerical analysis~\cite{Baker1975,Vidberg1977}.
From the dynamic susceptibility $\chi_{\rm sub}(\omega)$, we evaluate the reflection coefficient $r(\omega)$ using Eq.~(\ref{eq:r_def}).

\begin{figure}[tbp]
	\centering
	\includegraphics[width=8cm]{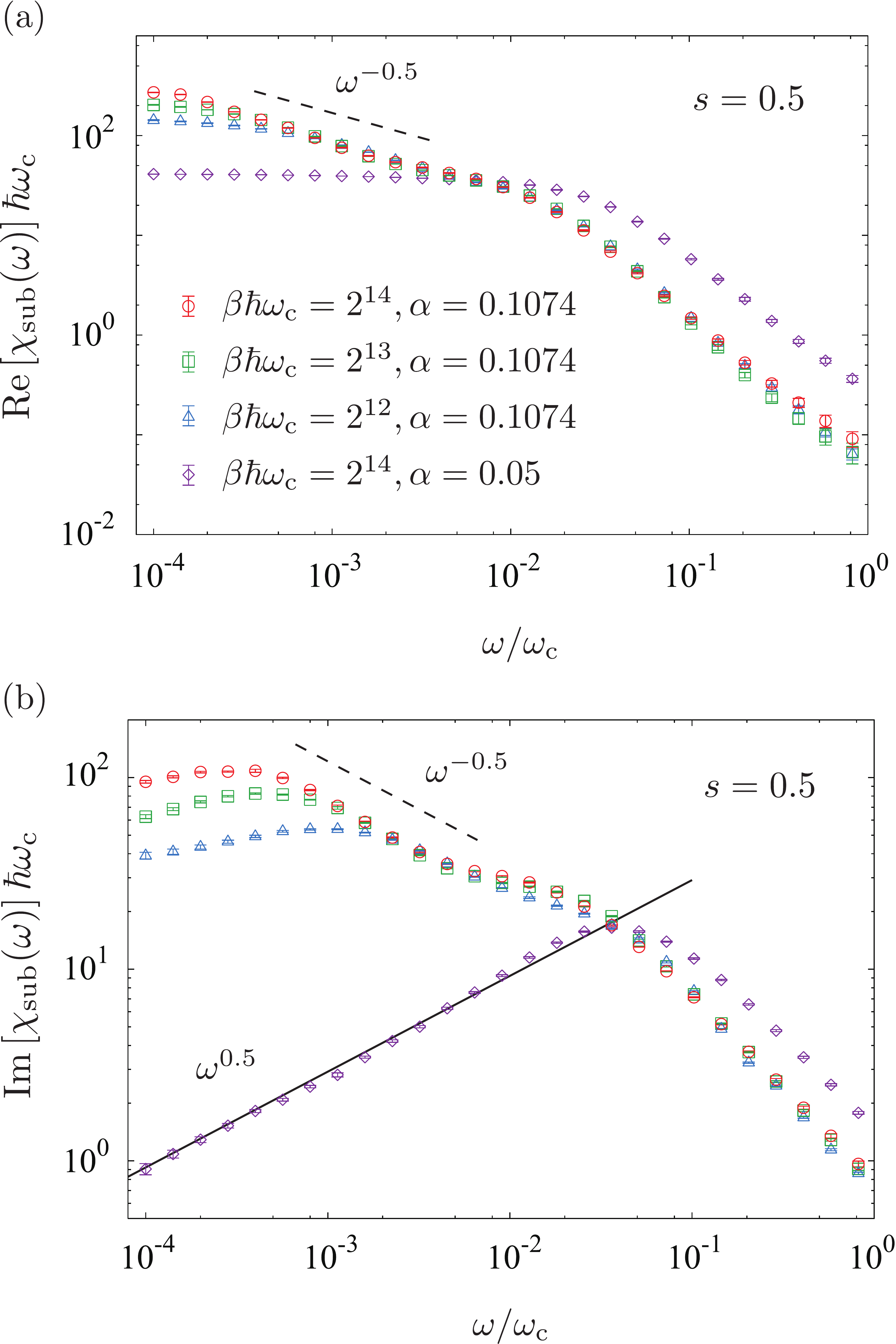}
	\caption{
    (Color online) Frequency dependence of the dynamic susceptibility for $s=0.5$ and $\Delta/\omega_{\rm c}=0.1$.
    (a) and (b) describe the plot of numerical results for the real and imaginary parts of the dynamic susceptibility calculated by the CTQMC simulations, respectively.
    The three plots indicate the results for $\alpha=\alpha_{\rm c}=0.1074$ at three different temperatures ($\beta\hbar\omega_{\rm c}=2^{12}$, $2^{13}$, and $2^{14}$), and the other plot for $\alpha = 0.05$ ($< \alpha_{\rm c}$) and $\beta\hbar\omega_{\rm c}=2^{14}$.
    The black solid line is given by the generalized Shiba relation~(\ref{eq:Shiba})}.
	\label{fig:dynamic_sus}
\end{figure}

\begin{figure}[tbp]
	\centering
	\includegraphics[width=8cm]{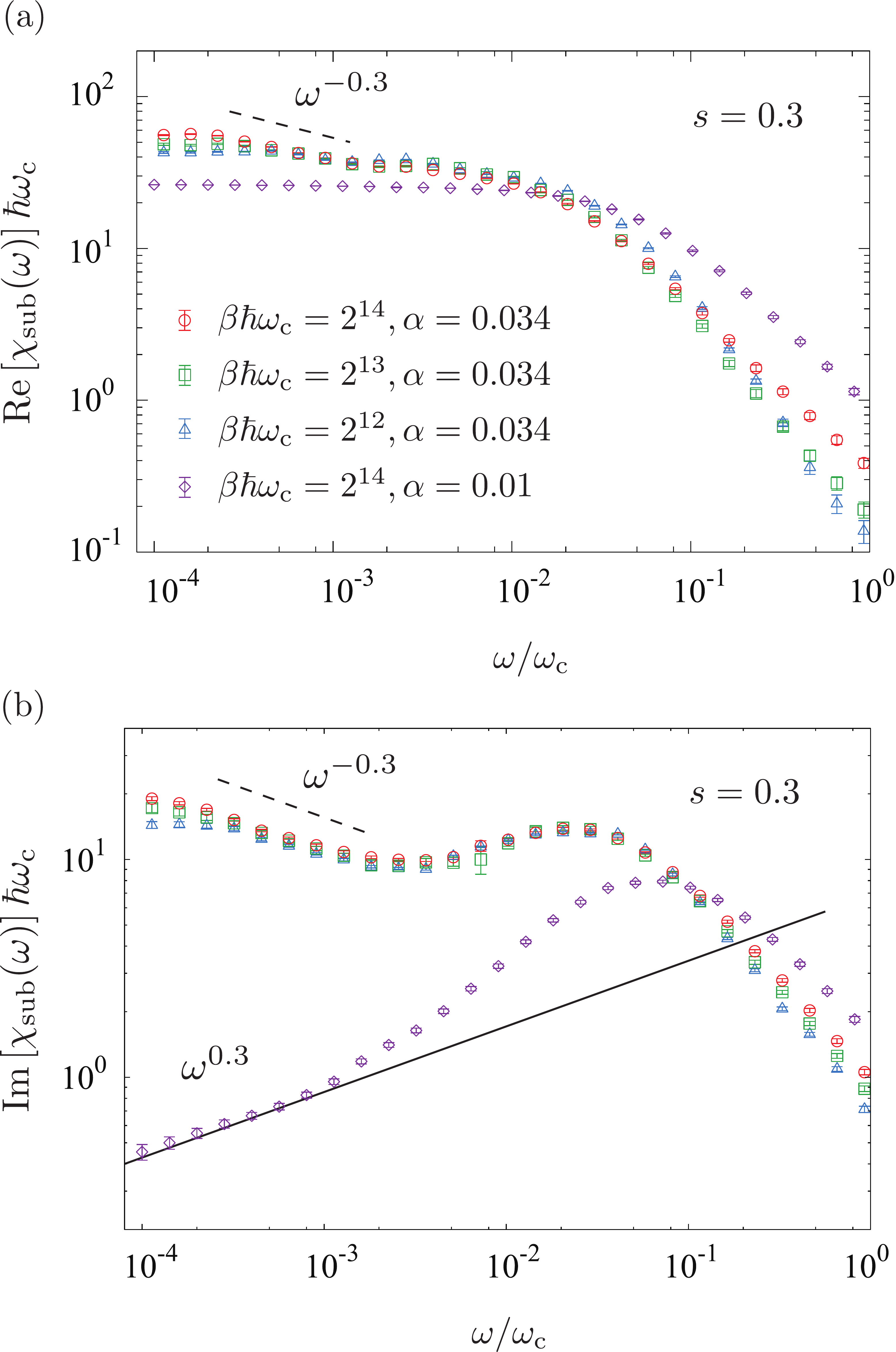}
	\caption{
    (Color online) (a) The real and (b) imaginary parts of the dynamic susceptibility for $s=0.3$ and $\Delta/\omega_{\rm c}=0.1$ as a function of frequency.
    The three plots correspond to the numerical results for $\alpha=\alpha_{\rm c}=0.034$ at three different temperatures ($\beta\hbar\omega_{\rm c}=2^{12}$, $2^{13}$, and $2^{14}$), and the other plot for $\alpha = 0.01$ ($< \alpha_{\rm c}$) and $\beta\hbar\omega_{\rm c}=2^{14}$. The black solid line is given by the generalized Shiba relation~(\ref{eq:Shiba}).
    }
	\label{fig:dynamic_sus_s03}
\end{figure}

We first discuss the dynamic susceptibility in the quantum critical regime ($\alpha = \alpha_{\rm c}$).
For demonstration purposes, we consider the case of $s=0.5$ and $\Delta/\omega_{\rm c}=0.1$, for which the critical point is determined as $\alpha_{\rm c}=0.1074$ by the Binder analysis (see Ref.~\citen{Yamamoto2018}).
Figures~\ref{fig:dynamic_sus}~(a) and (b) describe the plots of the real and imaginary parts of the dynamic susceptibility obtained by the CTQMC simulations as a function of $\omega$ at $\alpha = \alpha_{\rm c}$ for three different temperatures ($\beta \hbar \omega_{\rm c} = 2^{12}$, $2^{13}$, and $2^{14}$).
The dynamic susceptibility exhibits quantum critical behavior, $\chi_{\rm sub}(\omega)\sim\omega^{-y}$, as discussed in Sect.~\ref{sec:dynamic_sus_QCP}.
As shown in Fig.~\ref{fig:dynamic_sus}, the dynamic susceptibility is proportional to $\omega^{-0.5}$ for a frequency range, $k_{\rm B}T/\hbar \ll \omega \ll \tilde{\Delta}$, where $\tilde{\Delta} \lesssim \Delta$ is the renormalized tunneling frequency, which is a function of $s$, $\Delta/\omega_{\rm c}$, and $\alpha$.
Note that our result is consistent with those of previous studies~\cite{Bulla2003,Anders2007}.
The result for the low-frequency region is in good agreement with the critical exponent in Eq.~(\ref{eq:dynamic_sus_exponent}) for $s=0.5$.

We next consider the delocalized regime ($\alpha<\alpha_{\rm c}$) in the case of $s=0.5$ and $\Delta/\omega_{\rm c}=0.1$.
In Figs.~\ref{fig:dynamic_sus}~(a) and (b), we also plot the dynamic susceptibility for the coupling strength $\alpha = 0.05$, which is sufficiently smaller than the critical point $\alpha_{\rm c}=0.1074$. 
Additionally, the inverse temperature is set as $\beta \hbar \omega_c = 2^{14}$.
The real (imaginary) part of the dynamic susceptibility has a shoulder (peak) at a slightly higher frequency than $\omega = \tilde{\Delta}$ for the quantum critical regime, indicating that the renormalization effect of $\tilde{\Delta}$ due to the subohmic reservoir becomes stronger as the coupling strength increases.
For $\omega \ll \tilde{\Delta}$, the numerical result for the imaginary part of the dynamic susceptibility is consistent with the power-law frequency behavior obtained from the generalized Shiba relation~(\ref{eq:Shiba}), which is proportional to $\omega^{0.5}$ for $s=0.5$.

The same feature can be observed for other values of $s$. In Fig.~\ref{fig:dynamic_sus_s03}~(a) and (b), we show the real and imaginary parts of the dynamic susceptibility respectively in the case of $s=0.3$, for which the critical point is determined as $\alpha_{\rm c}=0.034$.
The other parameters are common as the calculation for $s=0.5$.
As expected, the dynamic susceptibility shows the critical behavior, $\chi_{\rm sub}(\omega)\propto\omega^{-s}$, at the quantum critical point ($\alpha=\alpha_{\rm c}$), although the frequency range showing the critical behavior is narrower than the case of $s=0.5$.
When the system is in the delocalized regime ($\alpha=0.01<\alpha_{\rm c}$), ${\rm Im}\left[\chi_{\rm sub}(\omega)\right]$ is proportional to $\omega^{s}$ as predicted from the generalized Shiba relation.

\begin{figure}[tbp]
	\centering
	\includegraphics[width=8cm]{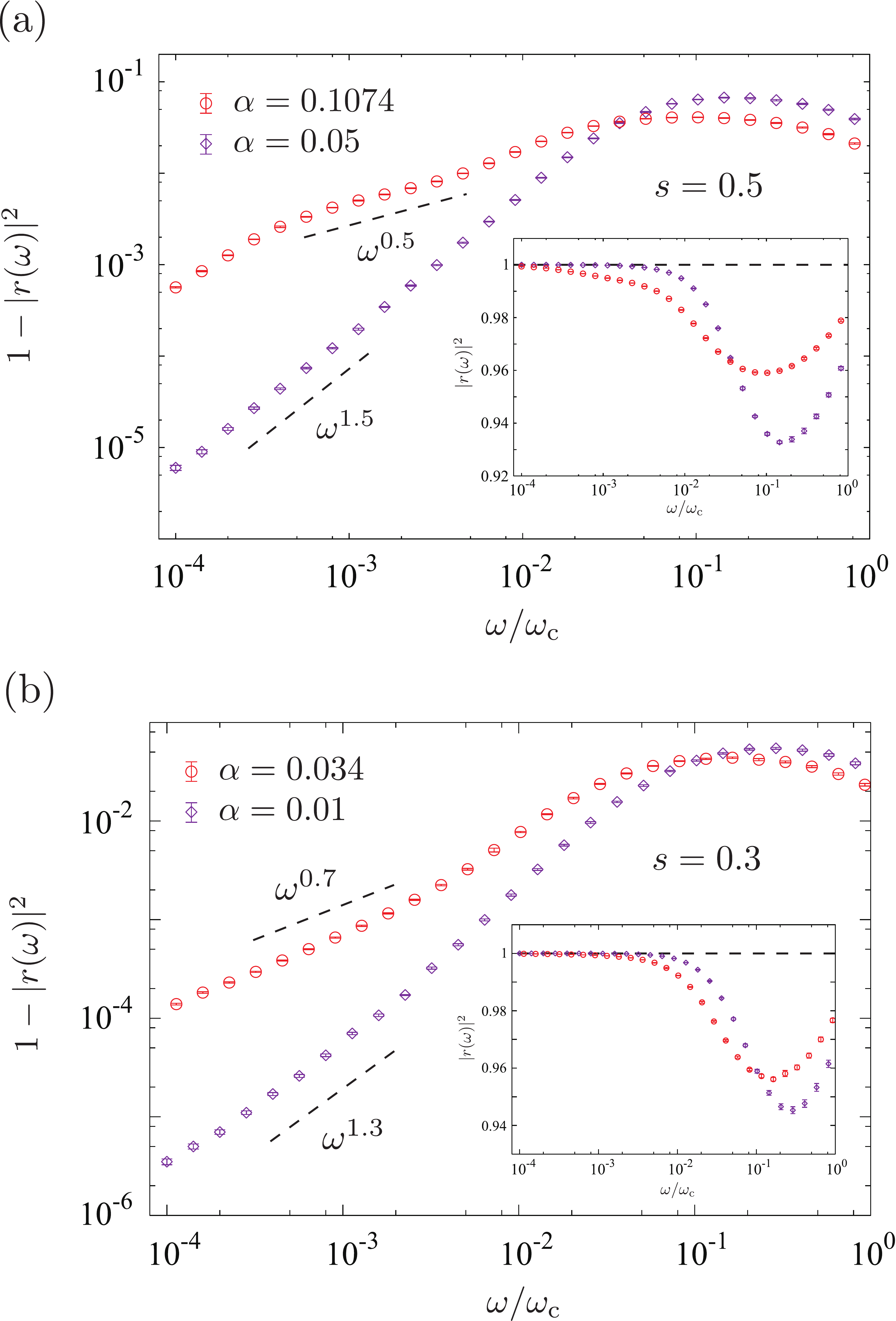}
	\caption{
    (Color online) Frequency dependence of the reflection calculated using the CTQMC simulations.
    The plots represent the numerical results for (a) $s=0.5$ and (b) $s=0.3$.
    The other parameters are set as $\Delta/\omega_{\rm c}=0.1$ and $\beta\hbar\omega_{\rm c}=2^{14}$.
    The red plots indicate the result for the critical point ($\alpha=0.1074$ for $s=0.5$, and $\alpha=0.034$ for $s=0.3$), whereas the purple plots indicate that for the delocalized regime ($\alpha=0.05$ for $s=0.5$ and $\alpha=0.01$ for $s=0.3$).
    }
	\label{fig:reflection}
\end{figure}

Figure~\ref{fig:reflection}~(a) and (b) depict the reflection loss, $1-|r(\omega)|^2$, for $s=0.5$ and $0.3$, respectively. 
The parameters are set as $\Delta/\omega_{\rm c}=0.1$ and $\alpha_{\rm line}=0.01$, the latter of which satisfies the weak coupling condition between the two-state system and the transmission line (the ohmic reservoir).
The reflection has a peak for both $\alpha = \alpha_{\rm c}$ and $\alpha < \alpha_{\rm c}$ for the renormalized tunneling amplitude $\tilde{\Delta}$.
For $\omega < \tilde{\Delta}$, the reflection loss approaches zero more slowly at the quantum critical point with a decrease in $\omega$.
From Eq.~(\ref{eq:r_def}), the reflection loss can be written as 
\begin{align}
    \label{eq:r_loss}
    1-|r(\omega)|^2
    = \pi\hbar I_{\rm line}(\omega){\rm Im}\left[\chi_{\rm sub}(\omega)\right] + \mathcal{O}(\alpha_{\rm line}^2).
\end{align}
Because $I_{\rm line}(\omega)\propto\omega$, the frequency dependencies of the dynamic susceptibility, as in Eqs.~(\ref{eq:dynamic_sus_QCP}) and (\ref{eq:dynamic_sus_co-tunneling}), would lead to reflection loss proportionality with $\omega^{1-s}$ in the quantum critical regime for $k_{\rm B}T/\hbar \ll \omega \ll \tilde{\Delta}$, and with $\omega^{1+s}$ in the delocalized regime for $\omega \ll \tilde{\Delta}$.
These frequency dependencies are consistent with the numerical result for $s=0.5$ and $0.3$ (see Fig.~\ref{fig:reflection}).

\section{Experimental Realization}
\label{sec:realization}

\begin{figure}[tbp]
	\centering
	\includegraphics[width=8cm]{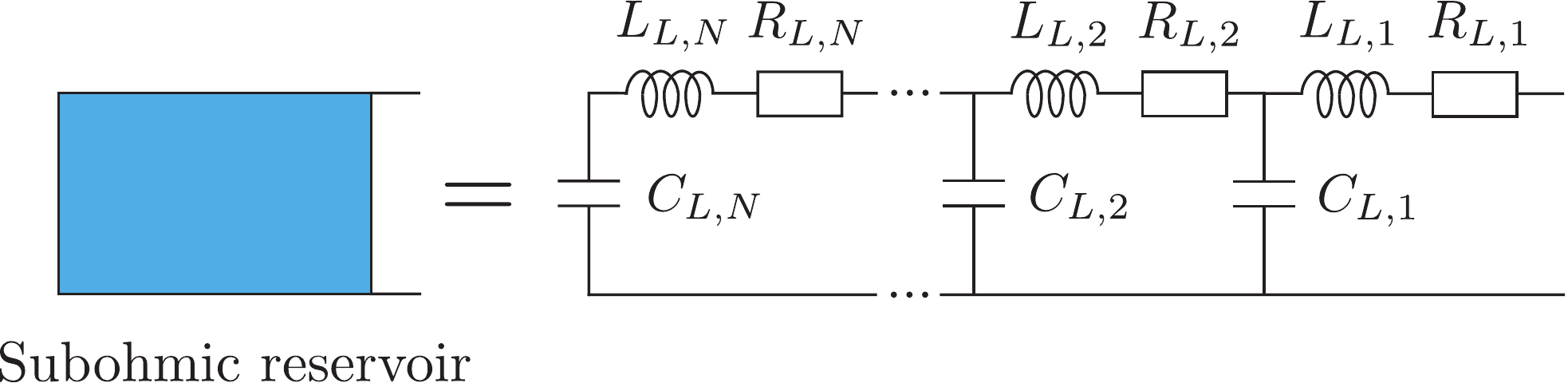}
	\caption{
	(Color online) The subohmic reservoir composed of an $RLC$ transmission line.
    }
	\label{fig:RLC_TL}
\end{figure}

In this section, we propose superconducting circuits, including $RLC$ transmission lines as shown in Fig.~\ref{fig:RLC_TL}, for the realization of the subohmic spin-boson model.
As previously stated in Sect.~\ref{sec:circuit_model}, the spectral density of the superconducting circuit is related to the total impedance as
\begin{align}
Z_{\rm sub}(\omega) = \frac{1}{Z_{L}(\omega)^{-1}+i\omega C_0},
\end{align}
where $Z_{L}(\omega)$ is the impedance of the $RLC$ circuit and $C_0$ is the capacitance connected to the output terminals in parallel (Fig.~\ref{fig:model} and \ref{fig:RLC_TL}).
Using Eq.~(\ref{eq:ImpedanceFormula}), the spectral density of the subohmic reservoir is
\begin{align}
    \label{eq:spectral_circuit}
    I_{\rm sub}(\omega) &= \frac{1}{\pi\hbar}\left(\frac{2|e|C_{J,L}}{C_{\Sigma}}\right)^{2}\tilde{I}_{\rm sub}(\omega), \\
    \label{eq:spectral_circuit_0}
    \tilde{I}_{\rm sub}(\omega) &= \omega{\rm Re}\left[Z_{\rm sub}(\omega)\right].
\end{align}
To obtain the impedance of the $RLC$ transmission line, $Z_{L}(\omega)$, we use the recurrence relation:  
\begin{align}
    Z_L(\omega) &\equiv Z_{L,1}(\omega), \\
    Z_{L,j}(\omega) &= R_{L,j} +i\omega L_{L,j} + \frac{1}{Z_{L,j+1}(\omega)^{-1}+i\omega C_{L,j}},
    \label{eq:RecurrenceRelation}
\end{align}
where $Z_{L,N+1}(\omega)^{-1}=0$ and $N$ is the number of repeated structures of circuit elements.

We first consider a simple circuit to realize the subohmic reservoir with $s=0.5$ in Sect.~\ref{subsec:s=0.5}.
We next expand the circuit model for an arbitrary value of $s$ smaller than $0.5$ in Sect.~\ref{subsec:s=0.25}.
Finally, we mention a circuit model for arbitrary $s$ in the range $0<s<1$.

\subsection{Subohmic reservoir of $s=0.5$}
\label{subsec:s=0.5}

\begin{figure}[tbp]
	\centering
	\includegraphics[width=8cm]{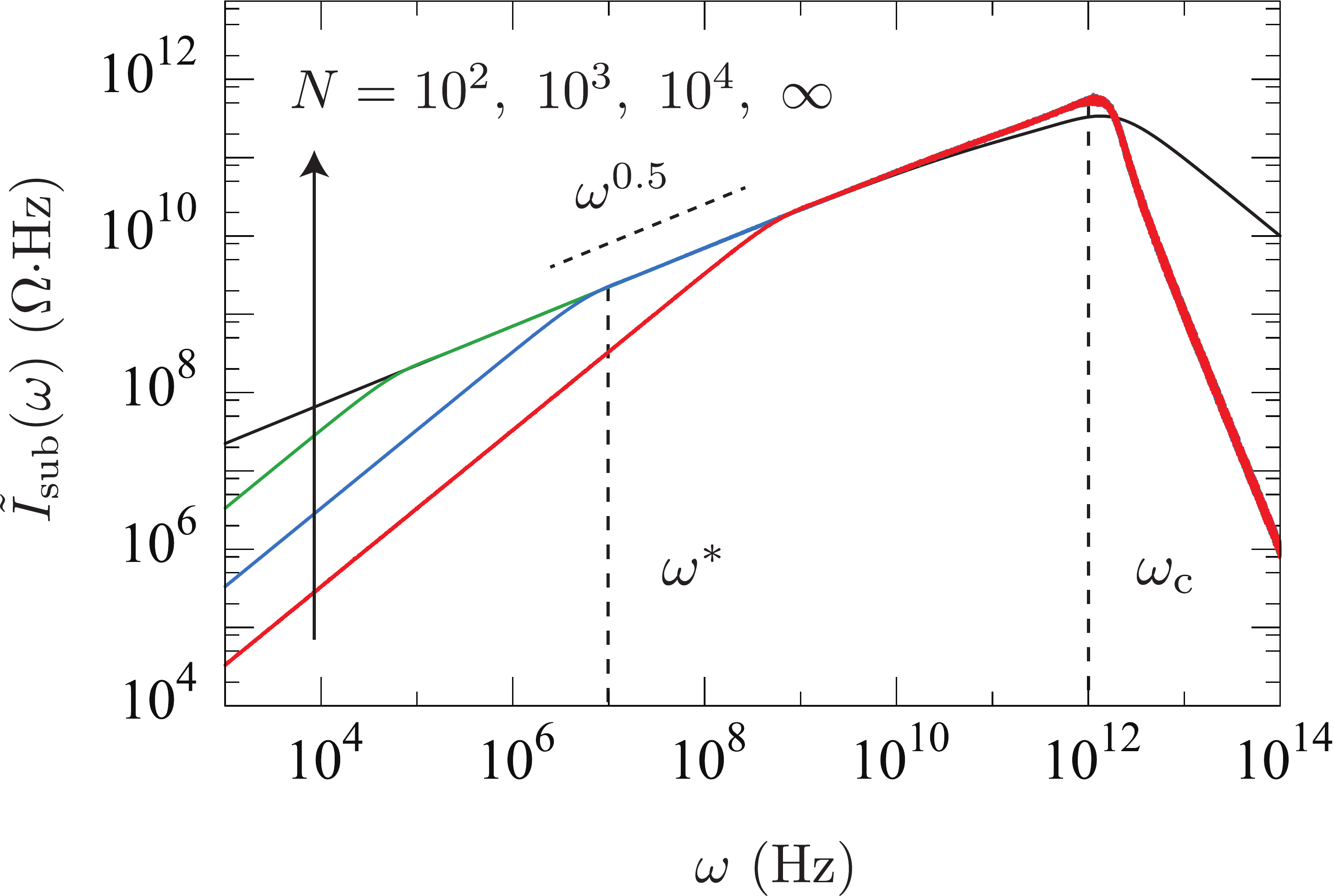}
	\caption{
    (Color online) Spectral density of the reservoir by the $RLC$ transmission line to realize the subohmic spin-boson system with $s=0.5$.
    The elements of the circuit are set as $R = 1~\Omega,~L=1~{\rm pH},~C=1~{\rm pF},~C_{0}=1~{\rm pF}$.
    }
	\label{fig:spectral_circuit_s05}
\end{figure}

Let us first consider a uniform $RLC$ circuit:
\begin{align}
R_{j}=R,\quad L_{j}=L, \quad C_{j}=C.
\end{align}
Figure~\ref{fig:spectral_circuit_s05} gives the spectral density, $\tilde{I}_{\rm sub}(\omega)$, of the $RLC$ circuit for $N=10^2,~10^3,~10^4$, and $\infty$.
Note that the case of $N=\infty$ has been discussed in Ref.~\citen{Tong2006}.
Here, the circuit parameters are set as $R = 1~\Omega$, $L=1~{\rm pH}$, $C=1~{\rm pF}$, and $C_{0}=1~{\rm pF}$.
To observe the effect of inhomogeneity in the circuit, we added $10\%$ relative randomness to the circuit parameters.
As shown in Fig.~\ref{fig:spectral_circuit_s05}, inhomogeneity in the circuit does not have any visible influence on the spectral density, except for the high-frequency region of $N=10^2$.
Figure~\ref{fig:spectral_circuit_s05} indicates that this circuit realizes the subohmic reservoir with $s=0.5$ for the frequency range, $\omega^{*} \ll \omega \ll \omega_{\rm c}$, where $\omega^{*}$ and $\omega_{\rm c}$ are the low- and high-frequency cutoffs, respectively. 
From the recurrence relation in Eq.~(\ref{eq:RecurrenceRelation}), these cutoffs can be defined as $\omega^{*}=\pi^2/(N^2RC)$ and $\omega_{\rm c}=C/(RC_{0}^2)$, respectively.
Fig.~\ref{fig:spectral_circuit_s05} also displays the low-frequency cutoff $\omega^*$ for $N=10^3$ and the high-frequency cutoff $\omega_{\rm c}$ (independent of $N$).
To realize the subohmic spin-boson system, we need to satisfy the condition $\omega^{*} \ll k_{\rm B}T/\hbar \ll \Delta \ll \omega_{\rm c}$.
Typical values of the tunneling frequency and the temperature for the charge qubit are $\Delta=35~{\rm GHz}$~\cite{Nakamura1999} and $k_{\rm B}T/\hbar=0.13~{\rm GHz}$, respectively~\cite{Magazzu2018}.
Therefore, the condition for observing QCP is well fulfilled for $N=10^3$.

\subsection{Subohmic reservoir of \ $0 < s < 0.5$}
\label{subsec:s=0.25}

\begin{figure}[tbp]
	\centering
	\includegraphics[width=8cm]{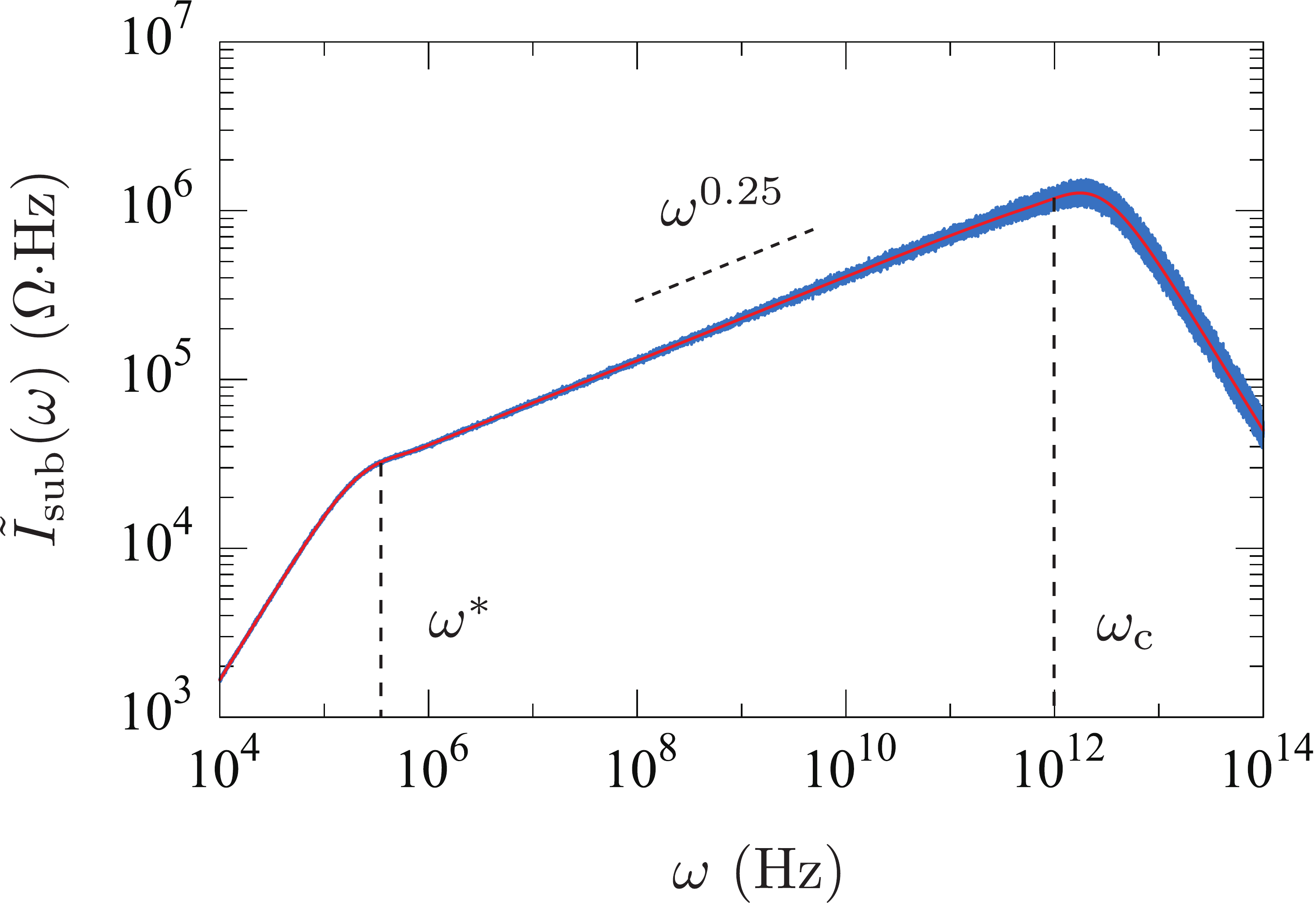}
	\caption{
    (Color online) Spectral density of the reservoir by the $RC$ transmission line to realize the subohmic spin-boson system of $s=0.25$ ($n=2$).
    The circuit parameters are $R = 50~{\rm m}\Omega,~L=0~{\rm H},~C=0.2~\mu{\rm F},~C_{0}=0~{\rm F},~N=10^2$.
    }
	\label{fig:spectral_circuit_s025}
\end{figure}

We next propose a circuit for the subohmic reservoir of arbitrary $s < 0.5$.
We consider an $RC$ circuit whose elements exhibit spatial dependence as
\begin{align}
    \label{eq:spatial_dependence_s025}
    R_{j} = R\left(\frac{j}{N}\right)^{n},\quad L_{j} = 0,\quad C_{j} = C.
\end{align}
Here, $n$ is a positive real number.
Figure~\ref{fig:spectral_circuit_s025} shows the spectral density, $\tilde{I}_{\rm sub}(\omega)$, for $n=2$ and $N=10^2$.
The circuit parameters are set as $R = 50~{\rm m}\Omega$, $L=0~{\rm H}$, $C=0.2~\mu{\rm F}$, and $C_{0}=0~{\rm F}$, to which we added $10\%$ relative randomness (blue line).
According to the figure, the present circuit realizes the subohmic reservoir for the frequency range of $\omega^* \ll \omega \ll \omega_{\rm c}$, where $\omega^*$ and $\omega_{\rm c}$ are the low- and high-frequency cutoffs, respectively.
Randomness in the circuit produces small fluctuations but does not change the overall feature of the spectral density for the case of no randomness (indicated by the red line). 
It is noteworthy that the subohmic reservoir can be realized for a wider frequency range extending to $N=10^2$.
We stress that the condition $\omega^{*} \ll k_{\rm B}T/\hbar \ll \Delta \ll \omega_{\rm c}$ is well fulfilled even for $N=10^2$, which is much less than in the case of the proposed circuit in Sect.~\ref{subsec:s=0.5}.

We can prove that for an arbitrary positive value of $n$, the spectral density is
\begin{align}
    \label{eq:spectral_circuit_s025}
    \tilde{I}_{\rm sub}(\omega) \propto \omega^{1/(n+2)},\quad (\omega^{*} \ll \omega \ll \omega_{\rm c}).
\end{align}
A detailed derivation is given in the Appendix~\ref{app:spectral_density2}.
This formula indicates that the present $RC$ transmission line potentially realizes the subohmic reservoir of $s < 0.5$ because $n$ is a real positive number.
This formula is consistent with Fig.~\ref{fig:spectral_circuit_s025}, in which the $RC$ circuit of $n=2$ realizes the subohmic reservoir of $s=1/(n+2)=0.25$.
The low- and high-frequency cutoffs are given by 
\begin{align}
    \label{eq:w^*_s025}
    \omega^{*} &= \frac{1}{RC}\left(\frac{n}{2N}\right)^2\left(1+\frac{2\sqrt{2}}{n}\right)^{n+2}, \\
    \label{eq:w_c_s025}
    \omega_{\rm c} &= \frac{1}{RC}\left(\frac{2N}{n}\right)^n,
\end{align}
respectively, and are shown in Fig.~\ref{fig:spectral_circuit_s025}.
Note that as $n$ decreases, the frequency range in which the spectral density behaves like the subohmic reservoir becomes narrower.

\subsection{Subohmic reservoir of \ $0 < s < 1$}
\label{subsec:s_arbitrary2}

Finally, we briefly discuss how to realize the subohmic spin-boson model with $0 < s < 1$, which includes the region of $0.5 < s < 1$ where the dynamic susceptibility, $\chi(\omega)$, has a nontrivial critical exponent in the quantum critical regime~(\ref{eq:dynamic_sus_exponent}).
We assume that the resistances and inductances depend on position with respect to
\begin{align}
    R_{j} = R\left(\frac{j}{N}\right)^{n},\quad L_{j} = L\left(\frac{j}{N}\right)^{p},\quad C_{j} = C,
\end{align}
where $n$ and $p$ are non-negative real numbers.
We can derive an analytic expression of the spectral density realized by this circuit as
\begin{align}
    \label{eq:spectral_circuit_arbitrary_s}
    \tilde{I}_{\rm sub}(\omega) \propto \omega^{2/(p+2)},\quad (\omega^{*} \ll \omega \ll \omega_{\rm c}),
\end{align}
where the low-frequency cutoff, $\omega^{*}$, is give by 
\begin{align}
    \label{eq:w^*_arbitrary_s}
    \omega^{*} = \left[\left(\frac{p}{2N}\right)^{2(n-p)}\frac{R^{p+2}}{L^{n+2}C^{n-p}}\right]^{1/(2n-p+2)},
\end{align}
and the high-frequency cutoff, $\omega_{\rm c}$, is a complex function of the element parameters in the $RLC$ transmission line.
Although this circuit can realize a subohmic reservoir of an arbitrary value of $s$ in the range of $0<s<1$, we need a very large number of circuit elements, $N\gtrsim10^{5}$, to achieve a sufficient frequency range where the spectral density behaves like the subohmic one.

\section{Summary}
\label{sec:summary}

We theoretically investigated QCP in the microwave scattering of a superconducting circuit.
We considered a system composed of a charge qubit and an $RLC$ circuit, which is effectively described by the subohmic spin-boson model, and studied the reflection coefficient of a microwave that is allowed to pass from a transmission line.
By performing the numerical calculation with the continuous-time Monte Carlo method, we clarified the frequency dependence of the dynamic susceptibility at and far from the QPT.
Moreover, we clarified how quantum critical behavior appears in the frequency dependence of a microwave loss.
We proposed three types of superconducting circuits to realize the subohmic spin-boson model and derived detailed conditions to observe QCP.
As such, this study is expected to provide an experimental platform to investigate QCP in a controlled manner.

{\footnotesize\paragraph{\footnotesize Acknowledgments}
The authors thank the Supercomputer Center, the Institute for Solid State Physics, the University of Tokyo for the use of the facilities.
TK was supported by JSPS Grants-in-Aid for Scientific Research (No. JP24540316 and No. JP26220711).
}

\appendix

\section{Analytic Expression of the Spectral Density}
\label{app:spectral_density2}

In this appendix, we derive the analytic form of the spectral density, Eq.~(\ref{eq:spectral_circuit_s025}), in the circuit model discussed in Sect.~\ref{subsec:s=0.25}.
Assuming $|\omega C_{L,j}Z_{L,j+1}(\omega)| \ll 1$, the recurrence relation~(\ref{eq:RecurrenceRelation}) can be rewritten into a differential equation in the limit $N\rightarrow \infty$: 
\begin{align}
    \label{eq:Riccati}
    \frac{\partial Z(\omega,x)}{\partial x} = -r(x)+i\omega l(x)+i\omega c(x)Z(\omega,x)^2, 
\end{align}
where $Z(\omega,x=j/N)\equiv Z_{L,j}(\omega)$, and $r(x)$, $l(x)$, and $c(x)$ are the resistance, capacitance, and inductance per unit length at $x=j/N$, respectively.
The total impedance of the $RLC$ circuit is obtained from the relation $Z_{L}(\omega)=Z(\omega,x\rightarrow0)$.
The spatial dependence of the circuit elements given in Eq.~(\ref{eq:spatial_dependence_s025}) can be rewritten as
\begin{align}
    r(x) = rx^n,\quad l(x)=0, \quad c(x)=c.
\end{align}
For $x \gg x^{*}\equiv (n^2/4\omega rc)^{1/(n+2)}$, the impedance is given by
\begin{align}
    Z_{A}(\omega,x) = \sqrt{\frac{r}{i\omega c}}x^{n/2},
\end{align}
due to the sufficiently small $\partial_{x} Z(\omega,x)$, in comparison with the other terms.
In contrast, for $x\ll x^*$, where the first term, $rx^n$, on the right-hand side of Eq.~(\ref{eq:Riccati}), can be neglected, the impedance is expressed as 
\begin{align}
    Z_{B}(\omega,x) = \frac{1}{-i\omega cx+A(\omega)},
\end{align}
where $A(\omega)$ is the constant of integration determined from the condition $Z_{A}(\omega,x^*)=Z_{B}(\omega,x^*)$.
Finally, we obtain the impedance of the $RLC$ circuit as
\begin{align}
    Z_{L}(\omega)
    &\sim Z_{B}(\omega,x\rightarrow0) \nonumber \\
    &= \frac{n}{\sqrt{2}\omega c}\left(\frac{4\omega rc}{n^2}\right)^{1/(n+2)}\left(1+i\frac{n+\sqrt{2}}{\sqrt{2}}\right)^{-1}.
\end{align}
for $\omega^{*} \ll \omega \ll \omega_{\rm c}$.
Therefore, the spectral density is obtained as
\begin{align}
    I_{\rm sub}(\omega) \propto \omega{\rm Re}[Z_{\rm sub}(\omega)] \sim \omega{\rm Re}[Z_{L}(\omega)]
    \propto \omega^{1/(n+2)}.
\end{align}
This expression shows the subohmic spectral density with $s<0.5$, which corresponds to Eq.~(\ref{eq:spatial_dependence_s025}).
Note that the high-frequency cutoff, $\omega_{\rm c}$, is obtained through the condition $\omega C |Z_{A}(\omega,x^{*})| \ll 1$, where the recurrence relation (\ref{eq:RecurrenceRelation}) can be reduced to the differential equation~(\ref{eq:Riccati}) for $x<x^{*}$.
The imaginary part of the impedance shows a sharp peak for $x \simeq 1$, which is neglected for $x>x^{*}$ in the above analysis, leading to the condition $\omega C(1-x^{*}){\rm Im}\left[Z_{A}(\omega,x^{*})\right] \gg 1$, by which the low-frequency cutoff $\omega^{*}$ is determined.

\bibliographystyle{jpsj_TY.bst}
\bibliography{Charge_qubit}

\end{document}